# Are Analogues of Hot Subdwarf Stars Responsible for the UVX Phenomenon in Galaxy Nuclei?

By B. DORMAN[1], R. W. O'CONNELL[1]
AND R. T. ROOD[1]

[1]Dept of Astronomy, University of Virginia, P.O.Box 3818 University Station, Charlottesville, VA 22903-0818 USA

We present the case that populations of sdB/sdOB/sdO-type stars may be a common constituent of galactic stellar populations, responsible for the UV upturn ('UVX') observed in the spectra of spiral bulges and normal galaxy nuclei. Extreme Horizontal Branch stars with $\log g > 5$ and $\log T_{\text{eff}} > 20,000 K$ have emerged in the last few years as the most likely candidate for the origin of the UVX. The magnitude of this far-UV flux in some systems (e.g. NGC 1399, NGC 4649) indicates that galactic nuclear regions must contain larger numbers of these subdwarfs than does the solar neighbourhood. This paper summarizes the results of a quantitative study of the UV radiation from evolved stellar populations. We have computed a large grid of stellar models in advanced stages of evolution, as well as a set of isochrones for ages 2-20 Gyr, for a wide range in composition. We use these calculations to derive synthetic UV colour indices for stellar populations with hot components.

We compare the results of this study to observations of the galaxies and clusters in the colours $m_\lambda(1500\text{Å} - V)$ $[= 15 - V]$ and $15 - 25$. In the globular clusters, the $(15 - V)$ colour is dependent on the size of the population of the blue horizontal branch (HB) and post-HB UV-bright stars. We assume that the magnitude of the UVX in the galaxies [as quantified by the $15 - V$ colour] is also dependent on the number of hot subdwarfs present. We thus determine the fraction of very blue stars that must be produced by an old metal-rich population in order to agree with the observations.

## 1. Introduction

The ultraviolet upturn ('UVX') observed in the spectra of elliptical galaxies and spiral bulges has emerged as an important issue in our understanding of stellar populations and the use of elliptical galaxies as tracers of the evolution of the universe. It was first recognized in observations of the M31 bulge from the OAO-2 satellite (Code 1969; see also Code & Welch 1979). Hills 1971 pointed out that the shape of this UV upturn was consistent with a hot thermal spectrum, and he suggested as the probable origin post asymptotic giant branch (P-AGB) stars, during their high luminosity ($\log L/L_\odot \sim 3-4$) transit to become white dwarfs. The UVX has since been found to be present in all elliptical galaxies, and to vary in amplitude by more than an order of magnitude. It has attracted considerable theoretical attention in recent years by researchers (see Greggio & Renzini [GR] and references therein) seeking to identify the hot component responsible for the UV radiation.

Candidate UV-bright stellar objects for this population include massive stars, very low-mass stars in advanced stages of stellar evolution, and low-mass products of binary interactions (see GR). Evidence against massive stars as the primary cause of the UVX has recently grown in strength (King *et al.* 1992; O'Connell *et al.* 1992). Further, Brocato *et al.* 1990 (see also GR) argued that populations of P-AGB stars are insufficiently long-lived to supply enough ultraviolet radiation to explain the strongest observed UV upturns. Currently, the most likely candidates are the Extreme Horizontal Branch (EHB) stars— stars that do not reach the thermally-pulsing stage of the AGB— and their post-HB descendants. These are termed Post-Early AGB or AGB-Manqué stars depending on whether they reach the lower AGB at all after core exhaustion. They have very thin hydrogen envelopes, and emit copious UV radiation both during and after the HB stage. In Dorman, Rood & O'Connell 1993, (hereafter DRO93) we presented a large grid of evolutionary tracks of these EHB stars with various metallicities and envelope masses. We use these here, together with isochrone computations, to model the UV colours of stellar populations containing hot components.

The grid of evolutionary tracks is defined by the composition parameters $Z$ and $Y$, the core mass $M_c^0$, and by $M_{env}^0$, the envelope mass of the object on the Zero Age Horizontal Branch (ZAHB). The parameter with the strongest influence on the integrated lifetime UV output of a star is $M_{env}^0$, whose distribution will be determined by the degree of mass loss on the red giant branch (RGB). The EHB stars are produced only if all but about 0.05 $M_\odot$ is lost at this stage of evolution. The empirically determined dependence of mass loss on both temperature and luminosity implies also that it should be a function of metal abundance. However, there is no suitable theory of the abundance dependence of mass loss (*cf.* GR). Since the physics here is so uncertain, our approach to the problem of the UVX is to concentrate on answering the following question: what ranges of $M_{env}^0$ are consistent with the observed UV fluxes of globular clusters and galaxies, given other constraints on their ages and abundances?

Our approach is to calculate directly the expected UV colours from evolved stellar populations by modelling the integrated flux from all phases of evolution. As a proving ground for the method, we consider the UV radiation from the Galactic globular clusters, showing the variation in UV properties that arises from different HB morphologies. We are able to demonstrate good agreement between the observations and synthetic colours in the colours $15 - V$, which measures the relative strength of the UV upturn — and $15 - 25$, which indicates the mean temperature of the UV-radiating stars in a stellar population. Turning to the galaxies, our principal result gives the fraction of the post-RGB stellar population that must radiate significantly in the UV in order to explain the range of $15 - V$ colours observed in the elliptical galaxies. Our principal conclusions are that (a) EHB stars are capable of producing more than enough far-UV flux to explain the observations, and (b) the resultant mid-UV flux (quantified by $15 - 25$) is also consistent with the data. Quantitatively, we find fractions of up to $\sim 20\%$ of red giants must become EHB stars after the helium core flash in the strongest UVX galaxies. Since the models we derive satisfy the available constraints provided by the data, they make a strong case for the hypothesis that EHB stars— which we identify with hot subdwarfs—are indeed responsible for the UVX.

## 2. The UVX Phenomenon in Galaxies

In this section, we briefly describe the observational characteristics of the UVX phenomenon. More details can be found in several recent reviews (Dorman, O'Connell, & Rood 1994, hereafter DOR94; O'Connell 1993; GR; Burstein et al. 1988, hereafter B3FL).
The features of the UVX phenomenon most germane to the question we pose here are:

• *Incidence:* It is found in the nuclear regions of almost all normal (i.e. excluding AGN and starburst) E's, S0's, and spiral bulges observed to date.

• *Strength:* The strength of the UVX, as measured by the colour $15 - V$, varies by about an order of magnitude, or 2.5 mag, (B3FL, and Figure 1). This degree of variation is much greater than that found in other optical and infrared broad-band colours.

• *Abundance Correlation:* As first pointed out by Faber (1983), and confirmed by B3FL, the magnitude of the UV upturn is positively correlated with the absorption line index Mg$_2$ (see Figure 1) in the *nuclear* regions of E/S0 galaxies and spiral galaxy bulges.

• *Radial Gradients:* Ultraviolet Imaging Telescope (UIT) data (O'Connell *et al.* 1992) show that M 31, M 81, and NGC 1399 have strong UV colour gradients with amplitudes up to $\sim 1$ mag in $15 - 25$, with the UV colours becoming strongly redder outward. In contrast, the dwarf elliptical M 32 has the reddest $15 - V$ in B3FL and a small UV colour gradient.

• *Resolution:* The more luminous UVX sources have apparently been resolved on recent HST Faint Object Camera exposures obtained of the M31 bulge. In a preliminary analysis of these data, King *et al.* 1992 detect $\sim$ 150 resolved objects, likely to be P-AGB stars and central stars of planetary nebulæ. However, these resolved sources account for about 17% of the total flux. Also, no massive OB stars appear to be in the field.

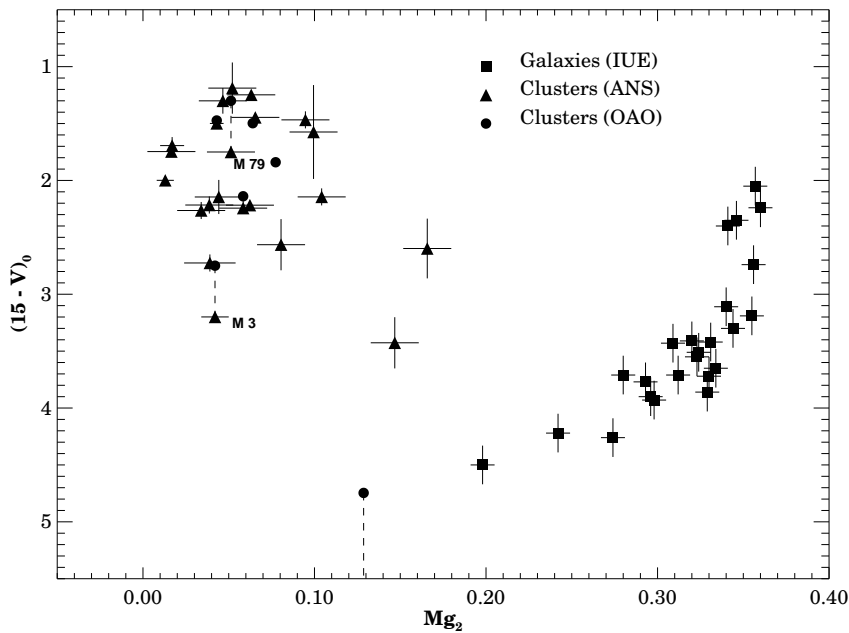

FIGURE 1. Observations of $15 - V$ colours of galaxies plotted against the absorption line index $Mg_2$. The galaxy data, derived from IUE SWP observations, are from B3FL, with the exception of NGC 1399 which is from Buson, Bertola, & Burstein (1993, private communication). The cluster data are from deBoer (1985). $Mg_2$ indices for the clusters are either observed or derived from the relation for Galactic globular clusters given by Brodie & Huchra (1990).

• *Spectral Shape and Features:* The IUE spectroscopy (B3FL) indicates that the far-UV spectral slope of the UVX is roughly constant from object to object and corresponds to $T_{\text{eff}} \gtrsim 20,000\,K$.

• *Temperature:* Far-UV (900–1800 Å) spectra of M31 and NGC 1399 were obtained by the Hopkins Ultraviolet Telescope (HUT) (Ferguson *et al.* 1991). The turnover of the spectrum at $\lambda < 1200$ Å indicates that $T_{\text{eff}} \lesssim 25,000\,K$.

• *Composite Nature:* The HUT spectrum of M 31 (Ferguson & Davidsen 1993) differs significantly from that of NGC 1399. For the adopted extinction, M 31 is hotter for $\lambda < 1200$ Å but cooler for longer wavelengths. Hence at least two distinct types of hot low mass stars produce the UV spectral energy distribution (SED) of M 31.

## 3. Ultraviolet-Bright Galactic Stellar Populations

We ask now, what are possible Galactic analogues of the UV-bright stellar population in the galaxies that could be responsible for the UVX, and which has characteristics that most closely resemble the galaxy spectra? We consider the UV flux from globular clusters and from the hot subdwarfs.

### 3.1. *Globular Clusters*

The globular clusters have long served as templates for evolved stellar populations, although their metallicity range does not strongly overlap that of the brighter galaxies (see Figs. 1 & 2). They do, however, they provide samples of homogeneous, coeval, old stellar populations which can be used to test synthetic UV colours. The clusters have been studied in the UV, from, amongst others, the ANS satellite (van Albada, deBoer, & Dickens 1981, hereafter ABD), OAO-2 (Welch & Code 1980, hereafter WC80), and Astro-1 (Hill *et al.* 1992; Whitney *et al.* 1994; Dixon *et al.* 1994). The far-UV flux from globulars is dominated by the hot HB stars, and is correlated with the HB morphology (ABD; WC80).

Even in the clusters with the strongest far-UV radiation, the SED is flatter than in the galaxies. The hardest far-UV spectral energy distributions among the clusters are observed in those with extended blue HB tails: even these, however, contain only a small population of true EHB stars, being largely populated instead by blue HB stars ($B - V \approx 0$).

The most metal-rich clusters, are, however, faint in the UV because their HB morphology is dominated by or exclusively composed of cool stars close to the RGB. One of the principal difficulties with the hypothesis that EHB stars are the origin of the UVX has, in fact, been the HB morphology in metal-rich globular clusters. If the UVX phenomenon is due to EHB stars, why should it be that they are produced in small numbers in the metal-poor globular clusters, not at all in the metal-rich clusters, and then reappear in the galaxies as the source of the UVX?

### 3.2. *Hot Subdwarfs*

There is evidence from the Galactic field that systems more metal-rich than the globular clusters produce stars which undergo sufficient mass loss to make EHB stars. This is the observed population of hot subluminous stars (Heber 1992; Saffer 1991). More than 1200 of these stars have so far been discovered, of which the majority were found by the Palomar-Green (PG) survey (Green, Schmidt, & Liebert 1986). The sdB stars, together with somewhat hotter objects termed either sdOB or sdB-O, constitute 40% of the catalogued stars. These occupy a narrow range of surface gravities with $\log g \sim 5 - 5.5$ and $20,000 < T_{\text{eff}} < 40,000$ K. The stars denoted as sdO —13% of the sample—have surface temperatures extending to about 65,000 K. Their surface gravities indicate a wide range in luminosity. About 34% of the remainder of the catalogued sources consists of hot white dwarfs and extragalactic objects, leaving very few that are directly attributable to products of binary evolution.

The derived temperatures and gravities for the sdB/sdOB stars match the predicted parameters of EHB stars. Further, the narrowness of the range in surface gravities is consistent with the notion that they are drawn from a single mass stellar population of about 0.5 $M_\odot$. For the sdO stars, there are several subclasses that may emerge from different progenitors, including the products of binary mergers (Iben 1990). There are reasons to believe that this is not the origin in the majority of cases (Heber 1992). The temperature and gravity range of the sdO stars suggests that at least some of them are the AGB-Manqué progeny of EHB stars. A crude comparison of the relative lifetimes of EHB stars and their AGB-Manqué progeny predicts a ratio of about 1:5. In addition, in the theoretical tracks, the greater part of the AGB-manqué lifetime is also spent at temperatures lower than $\sim$ 60-65,000 K. The observed ratio sdO:sdB thus appears to be about 1:3, with a deduced temperature range consistent with the theoretical sequences. We thus tentatively conclude, following a similar remark by Heber (1992), that *the bulk of sdO stars are evolved sdB and sdOB stars*. Finally, the disk membership of sdB/sdOB and sdO stars makes it likely that they originate from a metal rich population.

If the hot subdwarfs are produced in numbers in the Galactic disk, they are likely to be a part of the normal evolved stellar population of any galaxy. We may also expect that bulge of our Galaxy has a UV upturn analogous to that seen in the bulges of M 31 and M 81. Estimates of the numbers of hot subdwarfs in the solar neighbourhood (Heber 1992; Saffer 1991) indicate that their integrated UV flux is expected to be much smaller than the observed nuclear fluxes from galactic nuclei. The production of much larger numbers of subdwarfs thus seems to be a bulge rather than disk phenomenon. Finally, given that the UV upturn increases in strength toward the center of three of the systems observed by UIT including two spiral bulges, then if stars similar to subdwarfs are indeed responsible for a putative UV upturn in our Galaxy, their space density should also show a gradient. This may be observable in data samples collected from future field surveys in the bulge 'windows'.

## 4. Ultraviolet Observations

We present the data for the clusters in the same colour metallicity planes as the galaxies, so that we may easily compare and contrast the different systems. Full descriptions and tabulations of the data appear in DOR94. We construct the colours $15 - V$, $25 - V$, and $15 - 25$ from the data collected in the ultraviolet and in the $V$ filter, given by deBoer 1985 and B3FL. The recent IUE observations of Rich, Minniti, & Liebert 1993 have also been included in the $15 - 25$ plane, derived from their Table 2.

Figures 1 and 2 exhibit extinction-corrected $15-V$, $15-25$ colours respectively against the $Mg_2$ index. We interpret the $(15 - V)$ colour index as an indicator of the 'specific UV luminosity', measuring directly the number fraction of hot stars in the emitting population. The $15 - 25$ colour crudely represents the mean temperature calculated from all stars radiating significantly at wavelengths shorter than 2800 Å. However, since the mid-UV flux from the main sequence near the turnoff and the subgiant branch (SGB) can be significant, the $15 - 25$ colour can only be understood with composite models that include both the earlier phases of evolution as well as the hot star contribution.

Among the galaxies, the specific UV flux is largest in the most metal-rich objects, while the most metal-rich clusters (with red HB morphology) are fainter in the UV than those with lower metallicity. The bluest $15 - V$ colours among the clusters, however, are bluer than the galaxies with the strongest UV upturns. These clusters all have "extended blue HB tails", with some fraction of the stars hotter than 12,000 K (e.g. M 13, M 79, NGC 6752). The HB of M 3 extends almost to the red giant branch, and even its hottest HB stars are much cooler than most of the HB stars of M 13, despite their near-identical metallicity (Kraft *et al.* 1993). Thus in M 13 and similar objects, the entire HB phase contributes to the UV flux. In contrast, M 3, in which only a fraction of the HB population contributes to the far-UV radiation, is 2 mag fainter in $15 - V$. The location of the clusters in Fig. 1 can thus be seen as a sequence in which the UV flux decreases with increasing metallicity (the 'first parameter' effect), upon which is superposed a large scatter due to variations in HB morphology at intermediate ([Fe/H] $\sim -1.5$) metallicities.

In the $15 - 25$ colour index, the galaxies bluest in $15 - V$ are much bluer than the clusters. The reason for this is twofold: first, the clusters are more metal-poor, so that the light from the turnoff and SGB strongly influences the spectral shape in the mid-UV by reducing the slope longward of 2000 Å. Second, at shorter wavelengths, at which the earlier phases are invisible, the cluster fluxes do not rise with decreasing wavelength. Thus the cluster SEDs appear to be dominated by cooler stars than in the galaxies.

The contrast between the galaxies' and the clusters' UV colours is brought into sharp focus by the two colour diagram of $15 - V$ vs. $15 - 25$ shown in Fig. 3. The galaxies and the clusters form two almost parallel sequences of points. The metallicity index in the galaxy sequence increases from left to right (the UVX-$Mg_2$ correlation), while in the clusters it increases from right to left. An increase (decrease) in the fraction of hot stars present mainly produces a right (left) shift in the location of an object in this diagram. Any explanation for the UVX must be able to reproduce this behavior.

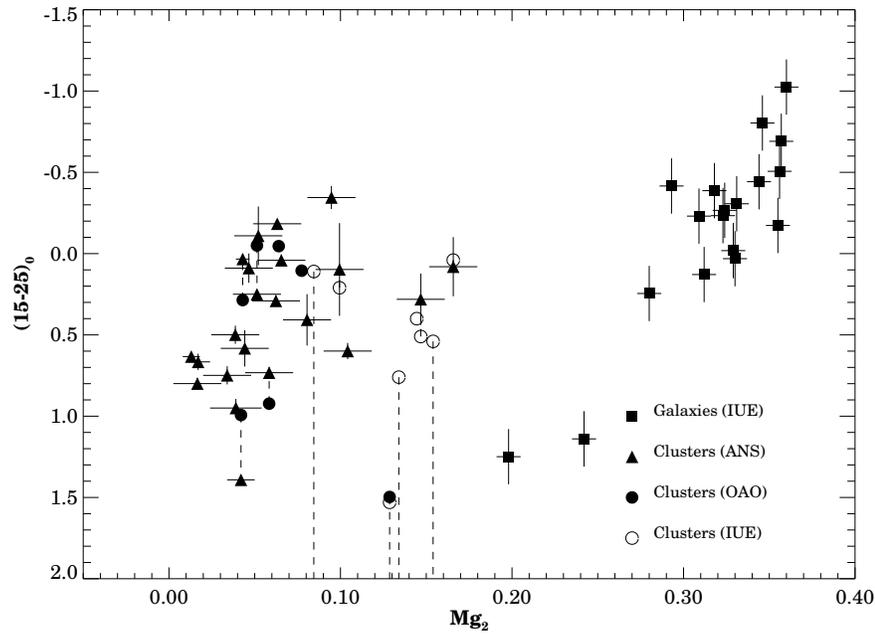

FIGURE 2. Observations of $15 - 25$ colours of galaxies plotted against the absorption line index Mg$_2$. The galaxy data are from B3FL, and derived from IUE SWP & LWP observations. The cluster data are from deBoer (1985), and Rich, Minniti, & Liebert (1993).

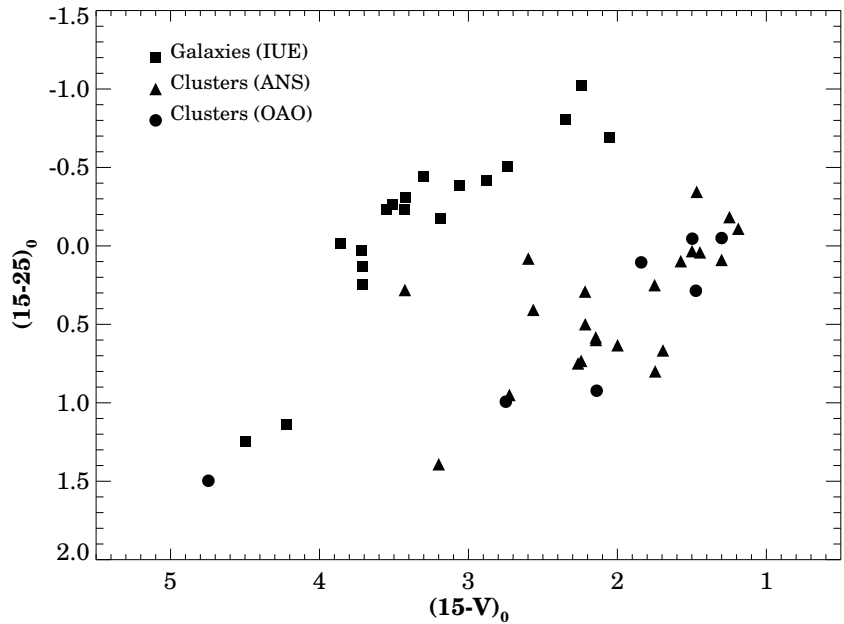

FIGURE 3. Observations of $15 - 25$ colours of galaxies plotted against the $15 - V$ colour. The data are the same as shown in Figures 1 & 2. The galaxies and clusters separate into two almost parallel sequences, with the abundance parameter increasing left to right in the upper sequence, and right to left in the lower.

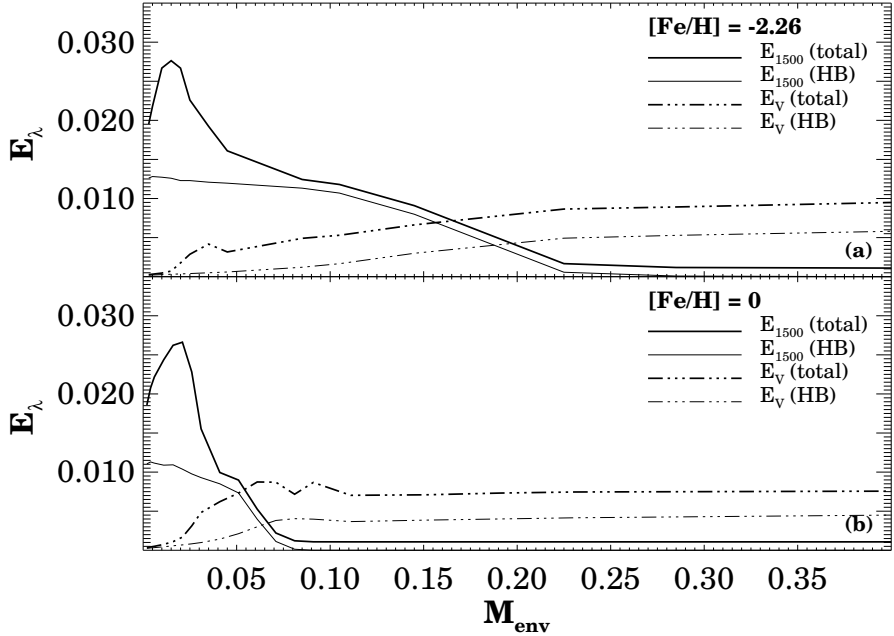

FIGURE 4. Integrated energy per unit wavelength radiated during HB & post-HB stages of evolution as a function of $M^0_{env}$, for the passband $1500 \pm 300$ Å and for the V-filter. The units of energy are $L_V^\odot$ Gyr Å$^{-1}$, and $M^0_{env}$ is in solar units. Panel (a) shows the behaviour for metal-poor HBs: note the range of masses with significant HB UV emission but little post-HB flux. Panel (b) shows the same for solar metallicity: here all models with large UV radiation are EHB stars.

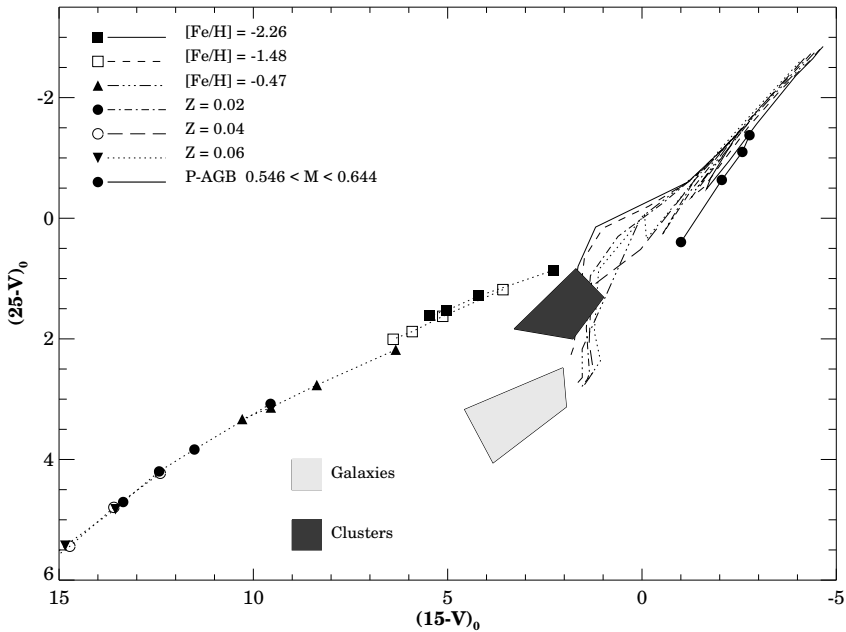

FIGURE 5. Synthetic colors from the various components of old stellar populations, plotted in mid-UV $(25 - V)$ vs far-UV $(15 - V)$ colours. The dotted lines with symbols show the colours of populations of ages 4-16 Gyr computed from isochrones and from the Salpeter mass function ($\Psi(m) = \mathcal{A}m^{-2.35}$). The upper right corner shows the location of 'pure' hot stellar components. A solid line connecting large filled circles shows the colours of P-AGB stars. The locations of the data of Figs. 1-3 are shown in the shaded areas.

## 5. Results

To calculate synthetic colours for composite stellar populations in the passbands of interest, we need (i) theoretical isochrones and luminosity functions for the phases up to the red giant branch tip (ii) the integrated energies emitted during the late evolutionary stages, and (iii) the 'specific evolutionary flux' of stars to the zero-age horizontal branch (ZAHB), i.e. the number of stars passing through the helium flash per unit $V$-luminosity of the population per unit time. A presentation of the necessary formulæ and a description of the calculations is given in DOR94. Briefly, the isochrones, combined with an assumption for the mass function, give the integrated colour from the earlier stages of evolution. It can be shown that the integrated flux along an HB/post-HB evolutionary track is equivalent, for a population of sufficient size, to the flux radiated by stars with the corresponding $M_{\rm env}^0$. The evolutionary flux is approximately equal to the rate of stars leaving the main sequence, which can be derived from a set of isochrones and the mass function at the turnoff.

Figure 4 shows the behaviour of the integrated flux with respect to $M_{\rm env}^0$ for two different compositions. It should be noted, first, that the maximum possible far-UV flux varies little with metallicity: thus any composition is capable of producing the observed UV colours. Second, for metal-poor compositions there is a range of mass that produces practically all of its UV flux during, rather than after, the HB phase. These stars evolve later to the AGB: in contrast, the 'true' EHB stars may produce as much UV radiation after core helium exhaustion as before. However, for metal-rich compositions (and this is true for all [Fe/H] $\gtrsim -0.5$), the only 'blue' HB stars are 'true' EHB stars.

Figure 5 shows the behaviour of individual components of an evolved population plotted in the $15 - V$ vs. $25 - V$ plane. The shaded areas mark the data plotted in Figs. 1-3. The dotted lines joined by symbols show the colours produced by the earlier phases alone, for ages 4-16 Gyr progressing from left to right. In the upper left of the diagram are curves representing the colours produced by 'pure' hot components, as a function of $M_{\rm env}^0$, for different metallicities. The solid line connecting filled symbols represents the range of flux produced by the Schönberner (1979, 1983) P-AGB model sequences.

The EHB stars appear uppermost and furthest to the right; as $M_{\rm env}^0$ increases, both colour indices move to the red. Once again, note that the colours of EHB stars of all compositions coincide, i.e. their UV flux is almost independent of metallicity. The opposite end of these curves represent the red HB stars whose far-UV flux is produced exclusively in the P-AGB phase. The $15 - V$ colours have been derived by adding the flux from the Schönberner 1983 lowest mass P-AGB model to our HB/AGB tracks. The $25 - V$ colour is, however, dominated by the stars in the HB phase and is unaffected by their final mass at the AGB tip. The red extremes plotted here represent the bluest possible colour that can be produced by P-AGB stars: if those present are more massive, the relatively cool, vertical end of the curves will move leftward.

For the comparison with the observations, note that:

(*a*) The colours of 'pure' hot components are much ($> 5$ mag) bluer than the observations. After adding the $V$-band contribution from the earlier stages ($\sim 20\%$ of the total $V$ flux), we can infer that the colours of the galaxies with the strongest upturns can be reproduced if a fraction $\ll 1$ of the red giants become EHB stars.

(*b*) The bluer $15 - V$ flux of the metal-poor clusters with blue HB tails implies a larger fraction of UV-emitting stars are produced in these systems.

(*c*) The $25 - V$ colours of the clusters are less than a magnitude bluer than the colours of the older metal-poor pre-HB component ($25 - V \sim 2$), implying that the earlier phases contribute significantly to this part of the spectrum.

(*d*) The galaxy colours are produced by the addition of flux from the 'isochrone' component, which for solar metallicity and above and at ages $\geqslant 10$ Gyr give $25 - V > 4$. The contribution at 2500 Å from the earlier phases is thus much smaller than in the clusters owing to the metallicity difference: the $25 - V$ colours of the *bluest* galaxies are dominated by the hot component (see B3FL).

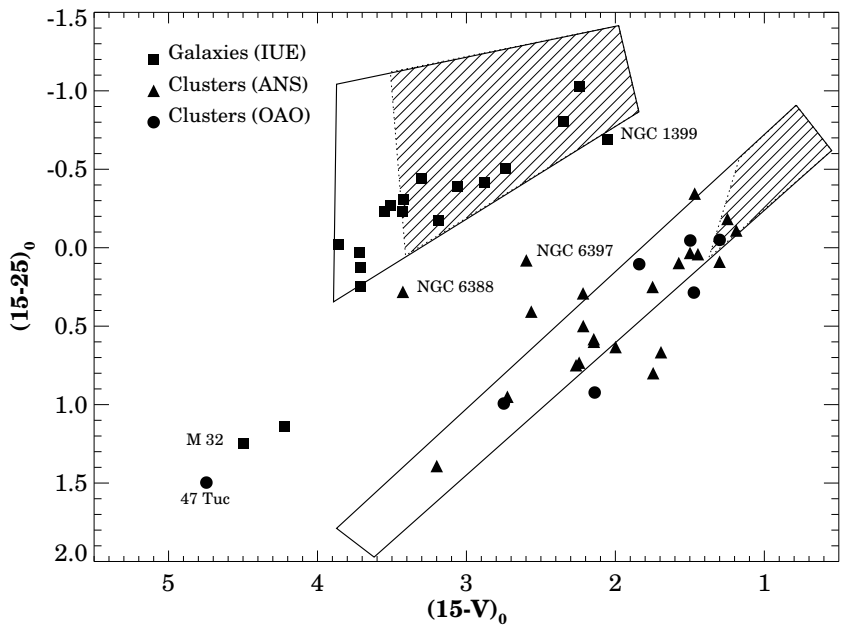

FIGURE 6. Results from simple models of the hot stellar component compared with the data of Fig.3. The upper box represents the location of models with $0 < [\text{Fe/H}] < 0.5$, $f_{UV} = 0.2$ (see text), at 10 Gyr. The shaded region indicates models with EHB stars and their post-HB progeny: the unshaded region consists of models with HB and P-AGB stars only. The lower box shows the location of models with $[\text{Fe/H}] < -1.5$, $f_{UV} = 0.75$, at 14 Gyr.

(*e*) The separation between the galaxies and the clusters on the two-colour diagram (Fig. 3) is induced by the effect of metallicity on the pre-HB component.

In general, both the number fraction of UV-bright stars and their envelope mass distribution will vary among systems. At high metallicity, the models contain only a small mass range at temperatures intermediate between the EHB and the red HB stars. In order to produce crude models of the populations containing a hot component, we introduce a simple parametrization of the HB mass distribution. The details of the models are not as important as the demonstration that they can reproduce the data using very simple, plausible assumptions. We divide the HB stars into two classes. The first, with fraction $f_{UV}$, varies in location on the HB between the EHB and the cool end. The other class consists of purely red HB stars, which, in the metal-rich case, all have similar output at 1500Å and $V$ to the reddest HB track (see Fig. 4b) and thus can be represented by a single sequence.

If we choose a value for $f_{UV}$ that reproduces the bluest galaxies, then add the flux from each HB track in turn to the radiation from the earlier phases of evolution, we can plot the result on the two-color diagram (Fig. 3). For $[\text{Fe/H}] \geqslant 0$ the result is the upper box in Figure 6. The shaded region in the box corresponds to those models in which the stars represented by $f_{UV}$ are EHB stars and their progeny. The unshaded region consists of models with HB and P-AGB stars only. In line with the expectation of (a) above and that in metal-rich populations the HB stars will be predominantly red, the plotted box has $f_{UV} = 0.2$ at 10 Gyr. Further, at the blue end the vertical spread is caused by metallicity, with the upper part of the box representing models with [Fe/H] $\sim 0.5$. However, the observed points lie in the lower half of the box, hinting that the true [Fe/H] of the galaxies may be close to solar (see Worthey, Faber & Gonzalez 1992). Note that the behaviour of the observations may also be reproduced by fixing the track representing the hot stars, and varying $f_{UV}$.

The lower box for the clusters has been generated similarly. In this case $f_{UV} = 0.75$, with the advanced phases added to a 14 Gyr population. We have thus fixed 25% of the stars to have similar flux to the red HB, and the other 75% are allowed to vary in mean location from blue to red. The models corresponding to the upper right of the box have very blue HB morphologies, while those at the lower left have purely red HBs. With this choice of $f_{UV}$ the colours of the clusters are fairly well reproduced, with only the bluest clusters falling into the edge of the shaded EHB area. This is consistent with (spectro-)photometric evidence that the clusters' UV flux is not dominated by EHB stars.

Finally, note that several of the outlying points are not inconsistent with the models. M 32 can be fit with models of age 4-6 Gyr, if P-AGB stars of slightly higher mass are used to provide the lower bound for the far-UV flux: specifically, if we use the Schönberner 0.565 $M_\odot$ model to approximate the far-UV flux of the red HB stars in the P-AGB phase. This estimate for the age of M 32, and the notion that its P-AGB stars are somewhat more luminous and massive, are both consistent with a growing body of similar findings (*cf.* O'Connell 1980; Freedman 1993). Apart from 47 Tuc whose far-UV flux is very small, the two most prominent clusters that lie away from the models are NGC 6388, which is fit by a model with [Fe/H] $\sim -0.5$, and NGC 6397, which has metallicity [Fe/H] $\sim -1$ and a bluer HB morphology. While we do not have models at this metallicity its location appears to be consistent with theoretical expectations.

It is a pleasure to thank Dave Burstein for providing data in advance of publication. We would also like to acknowledge useful conversations with Rex Saffer, Jim Liebert, & Pierre Bergeron. This research was supported by NASA Long Term Space Astrophysics Program grant NAGW-2596.